\documentclass[a4paper]{elsarticle}
\usepackage[latin1]{inputenc}
\usepackage{soul}
\usepackage{amsmath}
\usepackage{array}
\usepackage{latexsym}
\usepackage{geometry}
\usepackage{graphicx}
\usepackage{amsthm} 
\usepackage{amssymb}
\usepackage{float}
\usepackage{mathrsfs}
\title{Reducing local minima in fitness landscapes of parameter estimation by using piecewise evaluation and state estimation}
\author{Christoph Zimmer$^{*1}$, Frank T. Bergmann$^{2}$, Sven Sahle$^{3}$}
\begin{document}
%
%\linenumbers
\maketitle
\noindent\small
$ ^{1}$BIOMS, Im Neuenheimer Feld 267, 69120 Heidelberg, Germany\\
Email: christoph.zimmer@bioquant.uni-heidelberg.de;\\
$ ^{2}$BioQuant, Im Neuenheimer Feld 267, 69120 Heidelberg, Germany\\
Email: frank.bergmann@bioquant.uni-heidelberg.de;\\
$ ^{3}$BioQuant, Im Neuenheimer Feld 267, 69120 Heidelberg, Germany\\
Email: sven.sahle@bioquant.uni-heidelberg.de;\ \  \\
$ ^*$Corresponding author
\section*{Abstract}
\noindent
Ordinary differential equations (ODE) are widely used for modeling in Systems Biology. As most commonly only some of the kinetic parameters are measurable or precisely known, parameter estimation techniques are applied to parametrize the model to experimental data. A main challenge for the parameter estimation is the complexity of the parameter space, especially its high dimensionality and local minima. \\
Parameter estimation techniques consist of an objective function, measuring how well a certain parameter set describes the experimental data, and an optimization algorithm that optimizes this objective function. A lot of effort has been spent on developing highly sophisticated optimization algorithms to cope with the complexity in the parameter space, but surprisingly few articles address the influence of the objective function on the computational complexity in finding global optima. We extend a recently developed multiple shooting for stochastic systems (MSS) objective function for parameter estimation of stochastic models and apply it to parameter estimation of ODE models. This MSS objective function treats the intervals between measurement points separately. This separate treatment allows the ODE trajectory to stay closer to the data and we show that it reduces the complexity of the parameter space.\\
We use examples from Systems Biology, namely a Lotka-Volterra model, a FitzHugh-Nagumo oscillator and a Calcium oscillation model, to demonstrate the power of the MSS approach for reducing the complexity and the number of local minima in the parameter space. The approach is fully implemented in the COPASI software package and, therefore, easily accessible for a wide community of researchers.
%
%
%%%%%%%%%%%%%%%%%%%%%%%%%%%%%%%%%%%%%%%%%%%%%%%%%%%%%%%%%%%%%%%%%%%%%
%
\section*{Introduction}
\noindent
Mathematical modeling is a vital technique for analyzing complex systems in the sciences. We focus on models of ordinary differential equations (ODE) as these are widely used to model time course experiments in Systems Biology. These models are often parametrized using parameter estimation methods that find sets of parameters that let the model optimally fit to the data.\\[12pt]
Challenges arise from the complexity of the parameter space that may contain (many) local minima. Parameter estimation techniques consist of two components: an objective function quantifying how well the parametrized model fits the data and an optimization algorithm optimizing this objective function. A huge amount of research focuses on optimization algorithms: there are global optimization techniques \cite{Moles03,Rodriguez06,PS}, gradient based approaches \cite{Bock07, LM64} and Bayesian techniques \cite{Ghasemi11,Girolami08}. However, there have been few investigation on how the choice of the objective function leads to different landscapes influencing the difficulty of the optimization problem \cite{Bock07,Leander14}.\\[12pt]
We slightly extend a recently developed method \cite{Zimmer12} and show how it can be applied to ODE models to drastically reduce the complexity in the parameter search space. This ``multiple shooting for stochastic systems'' (MSS) method was recently developed for parameter estimation in stochastic models. The MSS approach splits the time course data into intervals that are treated separately. The initial values for each interval are composed from actual measurements and a state updating for the unobserved components. This article makes the approach more flexible by allowing to vary the number of intervals that are introduced. Using only one interval corresponds to the common least squares functional for ODE models. Using as many intervals as there is measurements corresponds to the MSS objective function in \cite{Zimmer12}.\\[12pt]
The importance of state estimation has been recognized in \cite{Bock07} that considers the state variables as optimization variables and uses continuity constraints on them to obtain a continuous trajectory in the end. Highly sophisticated structure exploitation reduces the complexity of the optimization problem again. \cite{Leander14} uses techniques from stochastic differential equations combined with a Kalman filtering. As both approaches and ours add flexibility to the states, this seems to be a crucial point in influencing the objective function landscape. The appeal of our method is the technical simplicity that gives any user the chance to easily implement it on their own. Even more, as the approach is implemented in COPASI, the full model import and analysis functionality of COPASI can be used.\\[12pt]
This article will demonstrate the benefits of using the extended approach to time course data from ODE models. The parameter estimation landscapes are a lot smoother with this MSS objective function which greatly simplifies optimization. It is fully implemented in the software COPASI \cite{Copasi} a popular modeling and simulation environment. Apart from its own file format, COPASI supports the import of models encoded in SBML \cite{hucka_2003full}, \cite{hucka_2010}. Therefore, a user can apply the method quickly to models available in model databases, or models created by hundreds of other SBML compliant software programs. \\[12pt]
The article will use three different example models to show the power of the MSS approach on ODE models: a FitzHugh-Nagumo oscillator, a Lotka-Volterra model and a Calcium oscillation model.
\section*{Method}
\subsection*{MSS objective function}
Assume that data $\nu^{\text{obs}} =(\nu_0^{\text{obs}},\ldots,\nu_n^{\text{obs}})$ is observed at time points $t_0,t_1,\ldots,t_n$. The state of the system at time $t_i$ is composed of the observed and unobserved species: $\nu_i=( \nu_i^{\text{obs}},\nu_i^{\text{hid}} )$. A model that describes the dynamical behavior of the system is given with a systems of ordinary differential equations:
\begin{gather}
 \label{eq:ode}
 \frac{d}{d t} x( t ; \theta ,x_0   )    =   f \left(t, x( t ; \theta, x_0  )  , \theta \right)\\
 x(0;\theta,x_0) = x_0,
\end{gather}
with a right hand side function $f$ depending on the systems state and some parameter $\theta$. In Systems Biology examples $f$ is often determined by the stoichiometric matrix $S$ and the rate law vector $v$: $f \left( x( t ; \theta, \nu_{i-1}  )  , \theta \right)=\ S\  v \left( x( t ; \theta, \nu_{i-1}  )  , \theta \right)$.\\[12pt]
We recently published an objective function for parameter estimation \cite{Zimmer12} that decomposes the whole time series into intervals and fits to the intervals individually. The initial value for each interval is obtained by a state updating. We suggest a slightly more flexible state updating here, namely the state updating is performed as follows: Choose a subset of the measurement time points: $T\subseteq \{t_0,t_1,\ldots,t_n\}$ and define the state update $\hat{\nu}_i$ at $t_i$ as:
\begin{equation}
\begin{aligned}
 \label{eq:state-update}
 \hat{\nu}_i^{\text{obs}} &= \nu_i^{\text{obs}},\ & t_i\in T  \\
 \hat{\nu}_i^{\text{obs}} &= x( \Delta_i , \theta ,\hat{\nu}_{i-1}  )^{\text{obs}},\ & t_i\notin T   \\
 &\text{ and } &  \\
 \hat{\nu}_i^{\text{hid}} &= x( \Delta_i , \theta ,\hat{\nu}_{i-1}  )^{\text{hid}}, & \text{ for all } t_i
 \end{aligned}
\end{equation}
with $\Delta_i= t_i-t_{i-1}$, ``obs'' denoting the observable components and ``hid'' the unobservable (hidden) components; hence the state estimate $\hat{\nu}_i$ is composed by $\hat{\nu}_i=( \hat{\nu}_i^{\text{obs}},\hat{\nu}_i^{\text{hid}} )$. The same holds for the ODE solution $x( \Delta_i , \theta ,\hat{\nu}_{i-1}  ) =  \left(  x( \Delta_i , \theta ,\hat{\nu}_{i-1}  )^{\text{obs}},     x( \Delta_i , \theta ,\hat{\nu}_{i-1}  )^{\text{hid}}    \right) $. \\[12pt]
The objective function for parameter estimation is then defined as
\begin{equation}
\begin{aligned}
 \label{eq:objfunc}
 F(\nu,\theta,\nu_0 ) &=   \sum_{i=1}^n  \left(  x(\Delta_i; \theta, \hat{\nu}_{i-1} ) - \nu_i \right) ^2 
 \end{aligned}
\end{equation}
with $x$ as in equation (\ref{eq:ode}), $\hat{\nu}_{i-1}$ as in equation (\ref{eq:state-update}). \\[12pt]
If $T=\emptyset$, the objective function equals the standard least squares functional (LSQ). If $T=\{t_0,t_1,\ldots,t_n\}$, then the objective function equals the MSS objective function in \cite{Zimmer12,Zimmer14}. The number of points in $T$ determines how many data points are used for updating the observable components of the system.\\[12pt]
%
%As comparison serves and ordinary least squares objective function that reads as:
%\begin{equation}
%\begin{aligned}
% \label{eq:lq}
% F_{\text{LSQ}}(\nu,\theta,\nu_0 ) &=   \sum_{i=1}^n  \left(  x( t_i; \theta, \nu_0 ) - \nu_i \right) ^2 \\
% \frac{d}{d t} x( t ; \theta ,\nu_0  )    &=   f \left( x( t ; \theta, \nu_0  )  , \theta \right), \\
% x(0;\theta,\nu_0)  &=  \nu_0 
% \end{aligned}
%\end{equation}
%Note, that the integration is here performed over the whole time horizon and the fitting is not split up into intervals.\\[12pt]
%
\subsection*{Optimization and Software}
The objective function can be optimized with gradient based methods, global 
optimization techniques or Bayesian approaches. The objective function is 
implemented in the software package COPASI. COPASI is a platform independent, user friendly software tool for 
setting up, simulating and analysing kinetic models of biochemical reaction networks.
It is freely available, open source software and is developed in a well established international
cooperation.
For a user of COPASI the workflow of running a parameter 
estimation with this objective function looks as follows: first the user would import
a dataset containing the observed/measured time points and map the columns of 
the dataset to the corresponding elements in his model. Once that is done, a feature 
in the \texttt{Tools} menu called \texttt{Create Events for Timeseries Experiment}
turns the measured time points into discrete events. These events force the value of 
a model variable at the measured time points to the observed value. Users are free
to modify the list of events, removing or adding further time points as desired. 
Removing all automatically created events restores the original LSQ functional. 
Example files have been placed on our website at \url{http://copasi.org/Projects/piecewise_parameter_fitting/}.

\begin{figure}[H]
  \label{fig:copasi}
   \includegraphics[width=\textwidth]{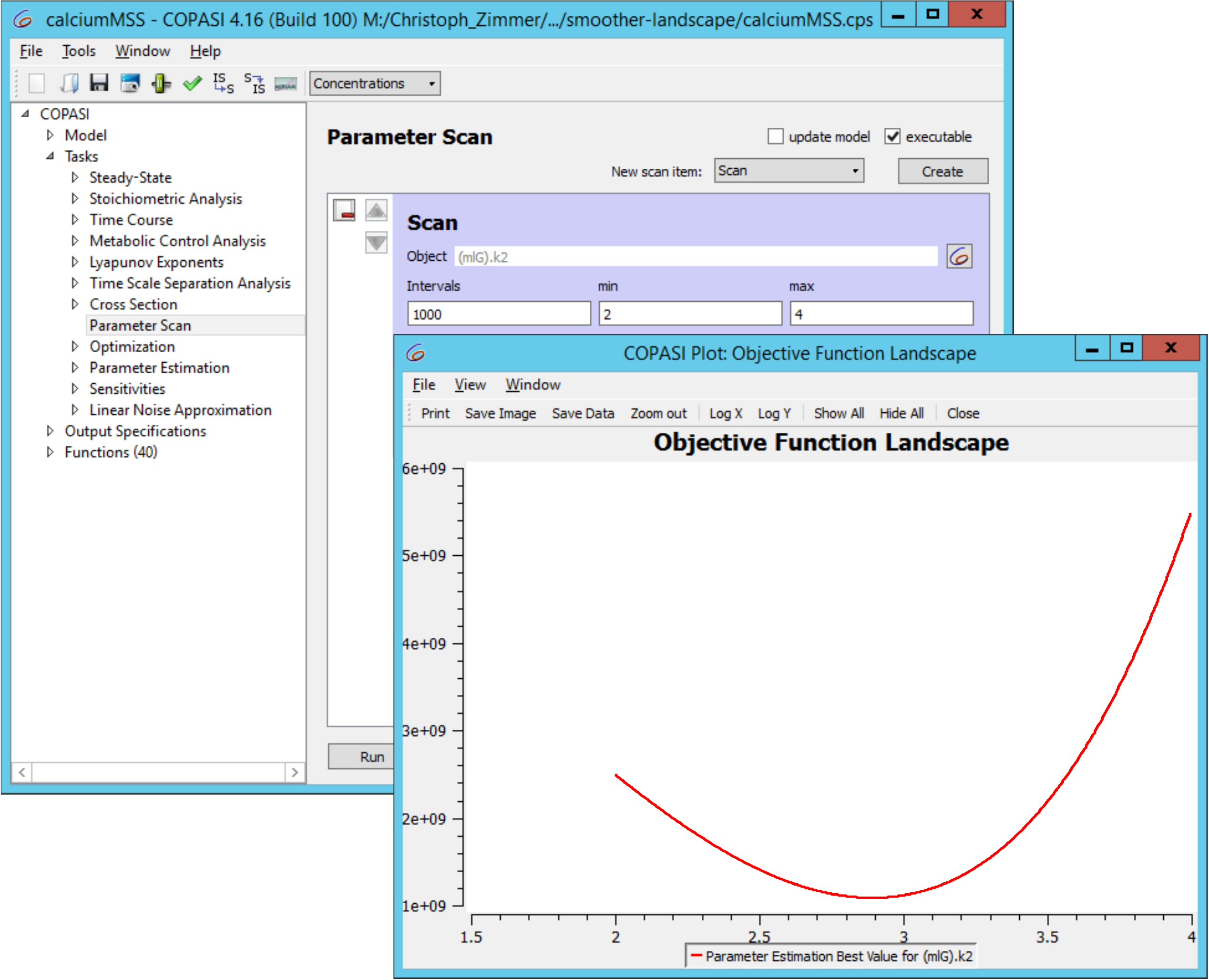}
   \caption{COPASI running a parameter scan plotting the objective function value for each parameter.}
\end{figure}
\newpage
\section*{Results}
The results section demonstrates the power of the method on three examples from Systems Biology: a FitzHugh-Nagumo oscillator, a Lotka-Volterra model and a Calcium oscillation model.
%
%
%%%%%%%%%%%%%%%%%%%%%%%%%%%%%%%%%%%%%%%%%%%%%%%%%%%%%%%%%%%% FitzHugh Nagumo
\subsection*{FitzHugh-Nagumo oscillator}
\begin{figure}[H]
    \begin{minipage}{1\linewidth}
   \begin{minipage}{.5\linewidth}
    \includegraphics[width=7cm,height=7cm]{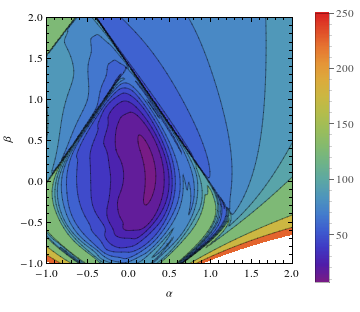}
  \end{minipage}
  \begin{minipage}{.5\linewidth}
    \includegraphics[width=7cm,height=7cm]{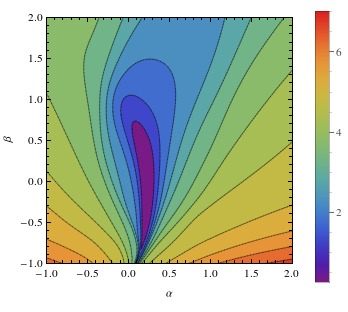}
  \end{minipage}
 \end{minipage}
 \caption{Fitness landscape of the FitzHugh Nagumo oscillator with LSQ objective function (left) and MSS objective function (right). Bright color stands for high values and dark color for small values. One can see that the LSQ objective function shows local minima whereas the MSS objective function is free of local minima.}
 \label{fig:fhn}
\end{figure}

The first example is a FitzHugh-Nagumo oscillator \cite{Nagumo62,FitzHugh61} with
\begin{eqnarray*}
\frac{dV}{dt} &=& \gamma \left( V-\frac{V^3}{3} + R \right)\\
\frac{dR}{dt} &=&  - \frac{1}{\gamma}\left( V-\alpha+\beta R \right)
\end{eqnarray*}
with initial values: $V(0)=-1$ and $R(0)=1$ and parameters $\theta=(\alpha,\beta,\gamma)$ with $\alpha=0.2$, $\beta=0.2$, $\gamma=3$. Measurements of $V$ are recorded at $t_k=0,1,2,\ldots,20$ with normally distributed additive noise with variance $0.1$. The parametrization has been chosen according to \cite{Leander14}.\\[12pt]
Figure \ref{fig:fhn} (left panel) shows the fitness landscape with the LSQ objective function ($K=\emptyset$) and figure \ref{fig:fhn} (right panel) shows the fitness landscape with MSS objective function ($T=\{0,1,2,\ldots,20\}$. One can see that the objective function landscape is a lot smoother using the MSS objective function rather than the conventional LSQ objective function. Therefore, optimization algorithms have less difficulties using the MSS objective function.\\[12pt]
As the MSS objective function differs from the LSQ objective function, their global minima are not necessarily identical. The LSQ minimum for this scenario is $(\alpha,\beta)=(0.25,0.19)$. Note, that the minimum is not the true parameter as measurement noise has been added to the pseudo data. The MSS minimum is $(\alpha,\beta)=(0.22,0.01)$. As most users will be interested in the LSQ minimum, one can use the MSS minimum as start value for a second optimization with the LSQ function. This second optimization will be very fast as the MSS minimum is within the valley of attraction of the global LSQ minimum. As described above both objective functions are implemented in the software COPASI \cite{Copasi}, making them readily available for users. They can enter their model using a graphical user interface, or start with 
models from model databases. Measured data is quickly applied to the model, and simulations can be easily carried out, plotted and analyzed. 
%
%
%%%%%%%%%%%%%%%%%%%%%%%%%%%%%%%%%%%%%%%%%%%%%%%%%%%%%%%%%%%% Lotka - Volterra
\newpage
\subsection*{Lotka-Volterra}
\begin{figure}[H]
  \begin{minipage}{1\linewidth}
   \begin{minipage}{.5\linewidth}
    \includegraphics[width=7cm,height=7cm]{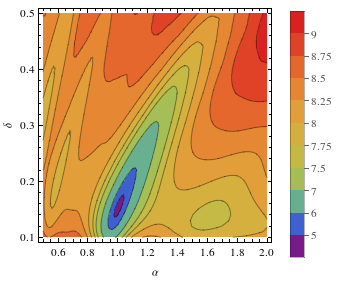}
  \end{minipage}
  \begin{minipage}{.5\linewidth}
    \includegraphics[width=7cm,height=7cm]{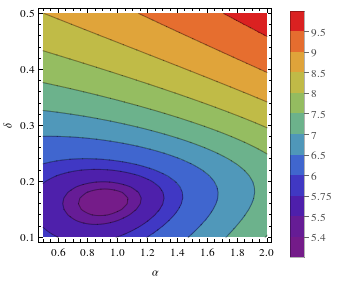}
  \end{minipage}
 \end{minipage}
 \caption{Fitness landscape for the Lotka-Volterra model with LSQ objective function (left) and MSS objective function (right). Bright color stands for high values and dark color for small values. One can see that the LSQ objective function shows local minima whereas the MSS objective function is smooth.}
   \label{fig:lv}
\end{figure}

The second example is a Lotka-Volterra system which describes the interaction of a prey population $X$ with a predator population $Y$:
\begin{gather*}
 \begin{aligned}
   \frac{d X}{dt}\ &=\ \alpha X - \beta X Y\\
   \frac{d Y}{dt}\ &=\ \delta X Y - \gamma Y.
 \end{aligned}
\end{gather*}
with initial values $X(0)=Y(0)=10$ and parameters $\theta=(\alpha,\beta,\gamma,\delta)$ with $\alpha=1$, $\beta=0.2$, $\gamma=1$, $\delta=0.15$. Measurements of $X$ and $Y$ are recorded at $t_k=0,0.5,1,\ldots,24$  with additive normally distributed measurement error with variance $1$.\\[12pt]
Figure \ref{fig:lv} (left) shows the objective function landscape with the LSQ objective function and figure \ref{fig:lv} (right) shows the objective function landscape using the MSS objective function. The first observation is that the MSS objective function landscape is free of local minima whereas the LSQ objective function landscape shows local minima. Therefore, optimization is much easier with the MSS objective function. \\[12pt]
Again, as the objective function are not identical, they lead to two slightly different minima: $(\alpha,\delta)=(1.01,0.15)$ for LSQ and $(\alpha,\delta)=(0.91,0.16)$ for MSS. However, the important point is that the MSS minimum is in the valley of attraction of the global LSQ minimum. This means that one can use the MSS minimum for a second optimization run with the LSQ objective function and this second optimization run is very fast.\\[12pt]
Furthermore, the Lotka-Volterra example has been used to investigate bias and variance of the LSQ estimates and MSS estimates. To this end, $100$ noisy time series have been created by adding gaussian noise with variance $1$ to a time course of $40$ observations with inter-sample distance $1$. Each of the $100$ time series is used for parameter estimation with the LSQ and MSS objective function, resulting in $100$ estimates $\hat{\theta}_k$, $k\in\{LSQ,MSS\}$ for each objective function. The average deviation of the estimates to the true value $\theta^{(0)}=(1,0.15)$ is calculated: This deviation $\frac{1}{100} \sum_{i=1}^{100} \left(  \hat{\theta}_k - \theta^{(0)} \right)$ is $(-0.11, 0.0010)$ for the MSS method and $(0.0034, 0.00065)$ for the LSQ approach. This means that the LSQ method has a smaller bias for this example. 
The standard deviation of the MSS estimates is with $(0.082, 0.011)$ larger than the standard deviation of the LSQ estimates with $(0.028, 0.0039)$.  
However, the authors suggest to use the MSS method for a first estimation and the MSS estimate as start value for a second LSQ estimation.\\[12pt]
\begin{figure}[H]
    \center{   \includegraphics[width=.67\textwidth]{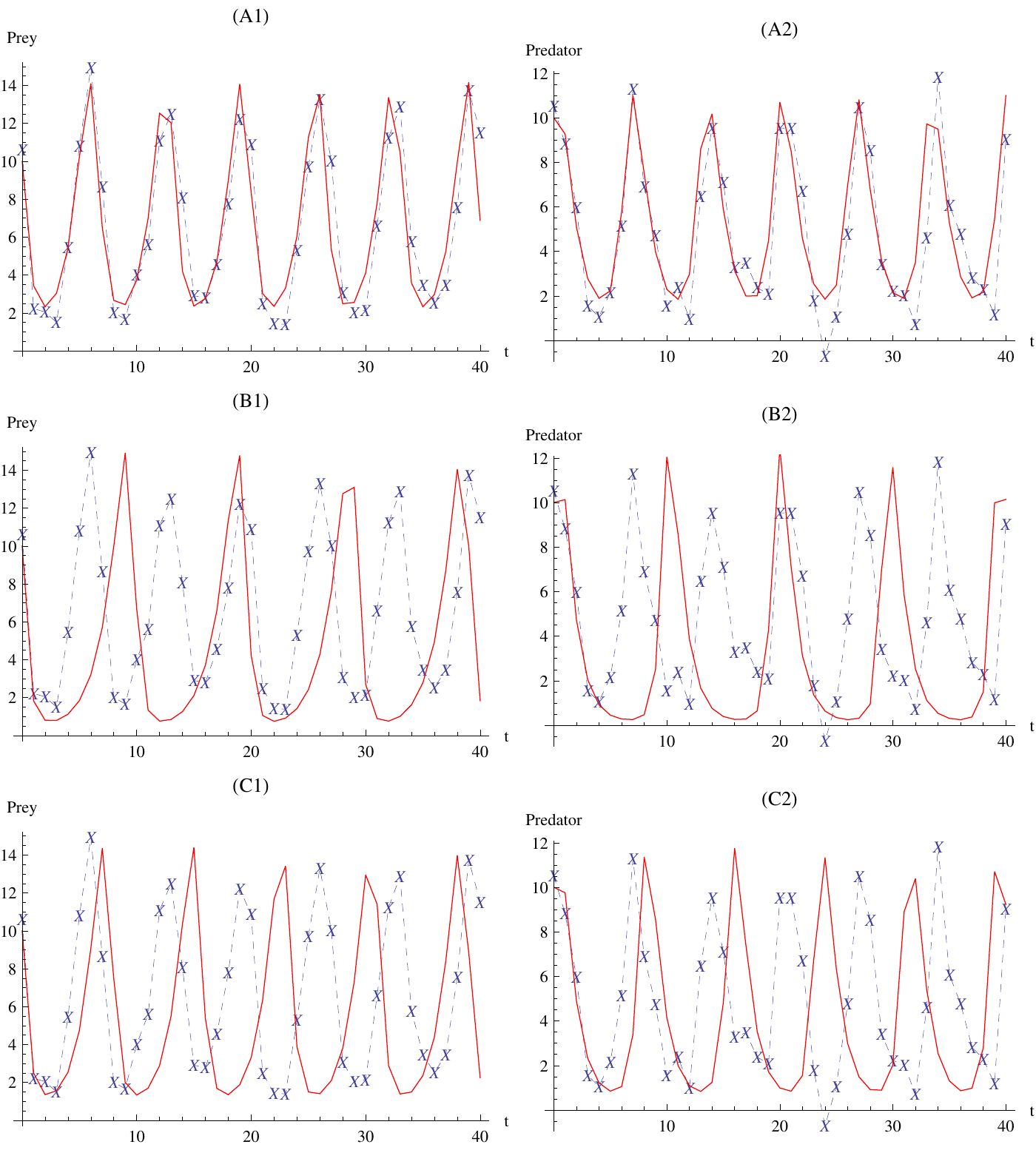}   }
 \caption{Showing fits with the LSQ function with (A1,A2) the global optimal parameter, (B1,B2) a local optimal parameter and (C1,C2) a parameter that is in the middle of global and local optimum.}
   \label{fig:lvlocmin}
\end{figure}
Figure \ref{fig:lvlocmin} illustrates why the LSQ objective function landscape has local minima: The global optimal fit in (A1,A2) shows good agreement of data and model. The fit in the local minimum (B1,B2) shows an oscillation with a different frequency, but still some of the peaks are fitted well. The parameter in the middle of global and local optimum, however, shows an oscillation that does not fit any of the peaks and results, therefore, in a worse objective function value. Our MSS method fits the data peace wise. This means that every interval is initialized with a state estimate that incorporates information from the observation. Therefore, effect as out-of-phase oscillation cannot occur any more. This seems to be a major effect in reducing the number of local minima in the objective function landscape.
%
%
%
%%%%%%%%%%%%%%%%%%%%%%%%%%%%%%%%%%%%%%%%%%%%%%%%%%%%%%%%%%%%%%%%%%%%%% Calcium 
\newpage
\subsection*{Calcium Oscillation model}
\begin{figure}[H]
    \includegraphics[width=\textwidth]{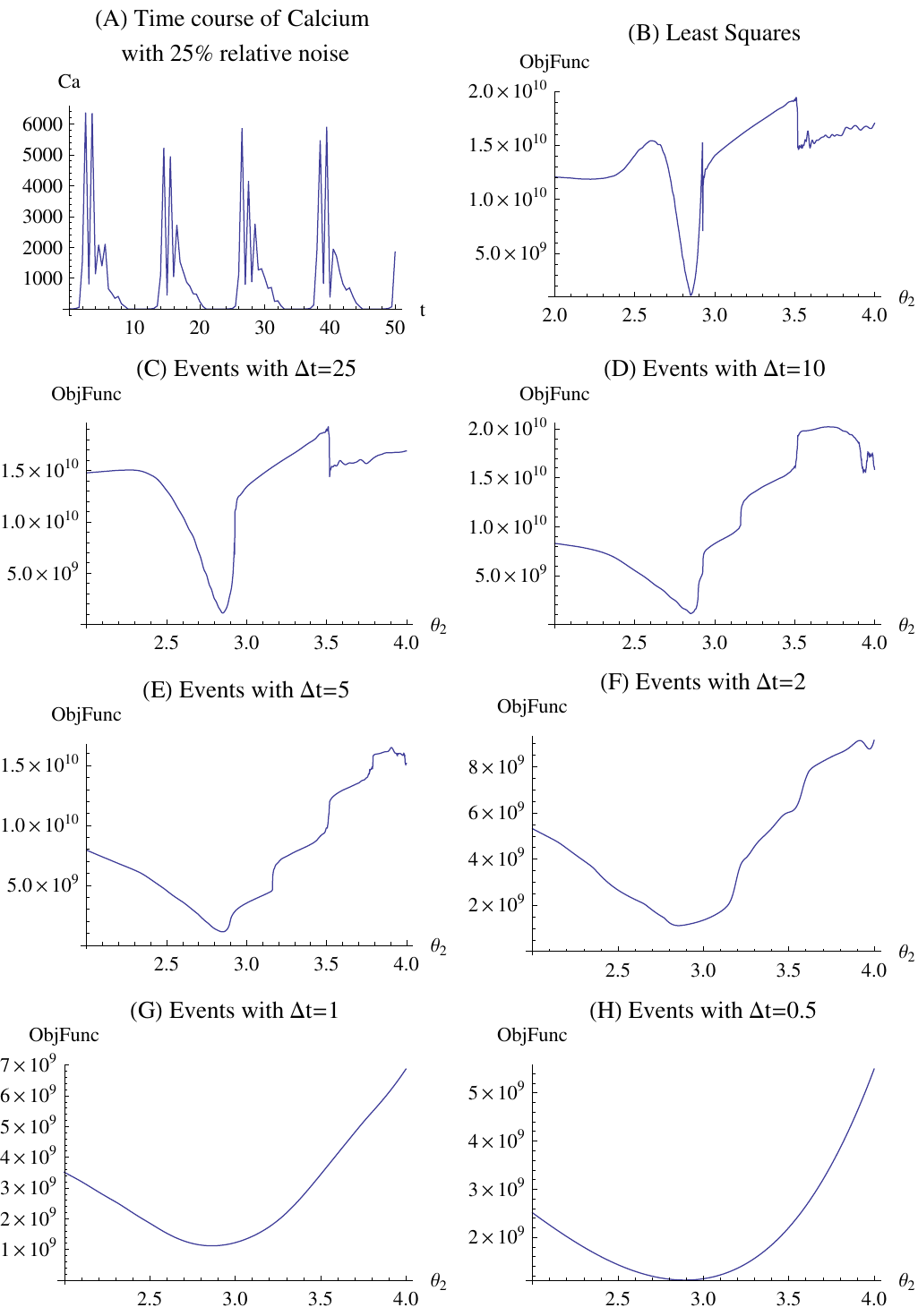}
   \caption{Calcium trajectory with 25\% relative measurement noise (A). Objective function landscape with LSQ (B) and the MSS (C-H) objective function. Panels C-H investigate the influence of the number of intermediate time points.}
   \label{fig:Ca}
\end{figure}

The third model is a Calcium oscillation core model from \cite{Kummer05}:
\begin{gather}
\begin{aligned}
\label{eq:ca}
  \frac{d g}{d t}\   &=\ \theta_1 + \theta_2 g - \frac{\theta_3\ g\ plc}{g+\theta_4} 
             - \frac{\theta_5\ g\ plc}{g+\theta_6}\\
  \frac{d plc}{d t}\ &=\ \theta_7 g - \frac{\theta_8\ plc}{plc+\theta_{9}} \\
  \frac{d ca}{d t}\  &=\ \theta_{10} g - \frac{\theta_{11}\ ca}{ca+\theta_{12}}. 
\end{aligned}
 \end{gather}
with initial values $g(0)=ca(0)=plc(0)=10$ and parameters \linebreak $\theta=(212, 2.85, 1.52, 190, 4.88, 1180, 1.24, 32240, 29090, 13.58,153000, 160)$. Measurements are assumed to be recorded at $t_i=0,0.5,1,1.5,\ldots,50$.\\[12pt]
The model with this parametrization exhibits a complex oscillatory behavior with bursting (figure \ref{fig:Ca} (A)), making parameter estimation especially challenging. 
Figure \ref{fig:Ca} (B) shows the LSQ objective function landscape when varying the parameter $\theta_2$. The landscape is full of local minima and optimization will require many function evaluations even on the relatively small interval $[2,4]$. Figure \ref{fig:Ca} (H) plots the MSS objective function landscape for the same setting. It is local minima free and convex which means that an optimization is simple. This shows again the beneficial influence of the MSS method on the objective function landscape. \\[12pt]
To address the question, how many intermediate points $T$ are necessary to reduce the influence of local minima, the objective function is also plotted with different values of $T$. This shows that already the introduction of only one intermediate point \ref{fig:Ca} (C) improves the landscape. However, local minima are still present and one needs, indeed, almost all observation points included in $T$ to obtain a local minima free landscape.\\[12pt]
If the MSS objective function is used with a nonempty $T$, it differs from the LSQ objective function which means that the minimum will not be necessarily identical. Estimating the parameter $\theta_2$ for all the scenarios one can see that the estimates are all in an interval from $[2.85,2.89]$. This means that the MSS objective function introduces 
only a small bias compared to the LSQ functional for this model. 
More importantly, it means that all the pre-estimates are in the valley of attraction of the global minimum of the LSQ functional. The software COPASI enables the user to use this pre-estimate as start value for a second run with the LSQ objective function. As this second run starts very close to the global minimum, an optimization is very fast.
\section*{Discussion}
\noindent
We extended a recently developed objective function for parameter estimation in stochastic models. This extension can be used for parameter estimation in models of ODEs and greatly simplifies the complexity of the parameter search space. We demonstrate this feature on three models where the LSQ objective function exhibits multiple local minina and show the power of our MSS objective function. \\[12pt]
The MSS method treats intervals between succeeding data points separately and uses a state updating to gain initial values for each interval. The extension increases the flexibility of this approach by allowing some of the intervals separately and some together. This extension allows to investigate the influence of state estimation on the objective function landscape.\\[12pt]
As \cite{Bock07} and \cite{Leander14} also add flexibility to the states, this seems to be a crucial point in influencing the objective function landscape. Figure \ref{fig:lvlocmin} illustrates how local minima can occur in the LSQ objective function landscape: As the LSQ approach uses only the very first initial value for integration, phase shifts in the oscillations might occur. This leads to an increase in the objective function (C1,C2). As soon as e.g. a double frequency is reached, the objective function value lowers again (B1,B2) leading to a local minimum. More generally, in complex systems an increase in deviation of the parameter from the true parameter does not necessarily lead to a stronger deviation of the model response from the data. However, approaches using state estimation reduce this problem as they are able to incorporate information from all observations (and not only the first) in the initial values for the intervals and, therefore, only require an integration over a shorter time interval. The special appeal of our method is the technical simplicity that gives any user the chance to easily implement it on their own. Even more, as the approach is implemented in COPASI, the full model import and analysis functionality of COPASI can be used.\\[12pt]
As the MSS objective function is not identical to the conventionally used least squares objective function, an MSS estimate will slightly differ from a LSQ estimate. This has been demonstrated for the Lotka-Volterra model for which the MSS method shows in a simulation study of $100$ pseudo data sets a slightly higher bias and variance. We do not have a general result but we think that the flatter curve of the MSS objective function leads to an easier optimization but as well an higher variance as experimental errors might shift the minimum of the curve stronger. Anyways, it is possible to use the MSS parameter estimate as already close start value for a (e.g. gradient based) LSQ based parameter estimation. As the computational cost of an optimization depends heavily on the landscape, these two optimizations have the advantage of both having local minima free landscapes. An additional benefit of the COPASI implementation of our approach is easy switching between the MSS and the LSQ optimization.\\
\section*{Acknowledgement}
\noindent
The software implementation of the presented method benefited from the efforts of the whole COPASI development team, currently Stefan Hoops, Brian Klahn, Ursula Kummer, Pedro Mendes, Ettore Murabito, and J\"urgen Pahle.
Christoph Zimmer is supported by BIOMS, Frank Bergmann by Virtual Liver \& NIH R01 GM070923 and Sven Sahle by the Klaus Tschira Stiftung. 
The funders had no role in study design, data collection and analysis, decision to publish, or preparation of the manuscript. 
\section*{References}
\bibliographystyle{unsrt}
\bibliography{120309}
%\bibliography{/home/cz/christoph-zimmer/LaTex/Literatur/bibtex/120309}
%
%
%
\end{document}